\documentclass[aps,prm,preprint,superscriptaddress]{revtex4-1}
\usepackage{braket}
\usepackage{amsmath}
\usepackage{graphicx}
\usepackage{color}

\begin{document}

\author{Sudipta Kundu}
\affiliation{Centre for Condensed Matter Theory, Department of Physics, Indian Institute of Science, Bangalore 560012, India}
\author{Mit H. Naik}
\affiliation{Centre for Condensed Matter Theory, Department of Physics, Indian Institute of Science, Bangalore 560012, India}
\affiliation{Department of Physics, University of California at Berkeley, Berkeley, California 94720, USA}
\affiliation{Material Sciences Division, Lawrence Berkeley National Laboratory, Berkeley, California 94720, USA }
\author{Manish Jain}
\email{mjain@iisc.ac.in}
\affiliation{Centre for Condensed Matter Theory, Department of Physics, Indian Institute of Science, Bangalore 560012, India}
\title
{Native Point Defects in Mono-- and Bi--layer Phosphorene}

\begin{abstract}
We study the stability and electronic properties
of intrinsic point defects, vacancy and self--interstitial,
in mono-- and bi--layer phosphorene.
We calculate the formation energies, quasiparticle defect states and
charge transition levels (CTLs) of these
defects using  \textit{ab initio} density functional theory (DFT) and GW
approximation to the electron self--energy.
Using the DFT + GW two paths formalism for studying interstitial in monolayer
phosphorene, we show that with the inclusion of electrostatic corrections 
CTLs can be calculated reliably.
Our calculations show that all the native point defects have low
formation energies 0.9--1.6 eV in neutral state.
Furthermore, we find that vacancy in phosphorene behaves as an acceptor--like
defect which can explain the p--type conductivity in
phosphorene. On the other hand, interstitial can
show both acceptor-- and donor--like behaviour.
\end{abstract}

\maketitle

\section{Introduction}
The discovery of graphene \cite{graphene1,graphene2} has evoked tremendous research in two--dimensional (2D) layered materials.
Apart from introducing new physics, 2D materials
show promising applications in opto--photonic, nano--photonic,
sensor devices etc \cite{graphene_book,graphene3}.
However, the semi--metallic nature of graphene limits its
wide application in electronic devices
despite its very high carrier mobility \cite{graphene4}.
Transition metal dichalcogenides \cite{mos21,mos22,mos24} are another important member of this 2D materials family.
Unlike graphene, they have a finite band gap, but their low mobility \cite{mos23} constrains their applications.
A few years ago, another 2D material, phosphorene, was exfoliated \cite{bP1,bP2,bP3,bP4} from bulk black phosphorus.
Phosphorene, the single or few layer form of black phosphorus, has been drawing much attention since then.
In a layer of phosphorene, a phosphorus atom is bonded to three
other phosphorus atoms via $sp^3$ hybridization.
This gives rise to a puckered honeycomb structure \cite{bP_struc}.
The uniqueness of phosphorene is its structural anisotropy
which manifests in anisotropic
optical, thermal and transport properties \cite{bP_aniso1,bP_aniso2,bP_aniso3}.
Phosphorene has a direct band gap at the center  of the Brillouin zone 
which ranges from $\sim$ 0.3 eV
(bulk) to $\sim$ 2.1 eV (monolayer) \cite{bP_expt2}. 
This change in the band gap with the number of layers is due to
quantum confinement and non-linearity of exchange--correlation
functional \cite{layer_dep}.
Phosphorene based field--effect transistors have shown a hole mobility of 1000 $\mathrm{cm}^{\mathrm{2}}/\mathrm{Vs}$ and a high on/off
current ratio up to $10^5$ \cite{bP1}.
The tunability of band gap and mobility depending on the number of layers make the material
suitable for electronic, opto--electronic, photo--electric, 
and FET device applications \cite{bP8,bP9}.

Presence of defect affects the material properties in different ways.
Depending on the charge state of the defect, it can induce free
carriers or trap charge and act as scatterer in the system.
Electronic, optical and magnetic properties can also be altered due to the presence of defects.
Hence, an extensive study of defects and their possible charge states is required for defect engineering and proper understanding of material properties.
The puckered structure of phosphorene facilitates the
formation of various defects \cite{vac1}. Defects in phosphorene have
low formation energy \cite{graphene_defect} in comparison to
defects in graphene \cite{bP_defect0}.
There is experimental evidence of vacancies \cite{bp_expt} in phosphorene.
While defects can escalate the degradation process of devices \cite{bP_oxy2,bP_oxy3}
such as vacancy can increase the oxidization of phosphorene, they can also enhance device
performance by introducing desired properties; for example, emission of
photons at new frequencies at room temperature \cite{bp_oxy}.
Vacancy and self--interstitial are two low energy native point defects in phosphorene.
In general, DFT works well for calculating ground state properties like structures of defects.
However, the estimation of the correct charge transition level (CTL)
requires the calculation of the energy associated with the change in the
electron number at the defect site. This is an excited state property and is not expected to be estimated correctly
with Kohn--Sham DFT.
While Heyd--Scuseria--Ernzerhof (HSE) functional \cite{hse} is an improvement over the generalized gradient approximation \cite{pbe}
and works well for bulk materials, it fails to capture the anisotropic screening
in two dimensional materials properly \cite{hse_prl,hse_prb}.
The combined formalism of many body perturbation theory within GW approximation \cite{gw_thry1,gw_thry2}
and DFT has emerged as a reliable method for calculation of CTL
\cite{gw_ctl,gw_ctl2}.
Furthermore, previous DFT calculations have either been performed with a small
supercell size \cite{bP_defect1} or have not taken into account the full
anisotropic (not only out-of-plane but also in-plane) dielectric constant
of phosphorene while correcting for spurious electrostatic interaction 
\cite{bP_defect2,bP_defect3,bP_defect4}.

We study structural and electronic properties of vacancy and self--interstitial in mono-- and bi--layer phosphorene.
We calculate the formation energies and CTLs 
of these defects using the DFT + GW formalism.
Since we employ periodic boundary conditions
in our calculations, the charged defect calculations suffer from
spurious electrostatic interactions with their periodic neighbours.
This problem is more prominent in a 2D material due to reduced screening along the out-of-plane direction.
We correct for this spurious interaction systematically by
properly modeling the anisotropic dielectric medium.
Further, we validate our calculation of CTLs by evaluating CTL of interstitial
in monolayer phosphorene following two different paths.
With the electrostatic correction to defect levels, the two paths agree within $\pm$ 100 meV.
We find that defects in phosphorene have low formation energies,
in the range 0.9--1.6 eV, which is consistent with
previous calculations \cite{vac1}.
Our study also shows that vacancy can induce
p-doping supporting the experimental finding of intrinsic p-type behaviour of phosphorene \cite{bP1, bP2, bP9}.
In contrast, interstitial in mono-- and bi--layer
can act as both p-type acceptor and n-type donor.
Furthermore, we find
that the unoccupied defect levels of neutral defect calculated with
DFT and GW line up with respect to the vacuum level
while the occupied defect levels shift
downward in GW compared to DFT calculated levels.
This manuscript is organized as follows. Section 2 describes our
methods of calculation which include the details about DFT, GW and the corrections for charged defect calculations.
Calculation and results of defects in mono-- and bi--layer phosphorene are
presented in section 3 and 4 respectively. 

\section{Computational Details and Methodology:}
All the DFT calculations in this study are performed
using the Quantum Espresso package \cite{qe}. We use a norm--conserving pseudopotential \cite{norm_pseudo}
and the exchange--correlation potential is approximated by the generalized
gradient approximation proposed by Perdew, Burke and Ernzerhof (GGA--PBE) \cite{pbe}.
The van der Waals interaction between layers is described by
\textquotedblleft Grimme--D2\textquotedblright{} method \cite{vdw}.
The wave functions are expanded in a plane wave basis, with plane waves up to energy  of 60 Rydberg included in the basis.
For unit cell
calculations we adopt a 14 $\times$ 10 $\times$ 1 Monkhorst--Pack \cite{monk-pack} k-point sampling of the Brillouin zone.
A vacuum space of 15.9 \AA{} is added to the unit cell in the out-of-plane
direction to simulate an isolated mono-- or bi--layer.
The in-plane lattice constants are found to be 3.30 \AA{} and 4.62 \AA{} for monolayer and 3.31 \AA{} and 4.51 \AA{}
for bilayer unit cell respectively. For the calculation of defects, we construct a 7 $\times$ 5 $\times$ 1
supercell to accommodate the vacancy or the interstitial atom.
This size of supercell is chosen to simulate an isolated defect and minimize the
interaction between periodic images of defects.
The Brillouin zone is sampled with a 2 $\times$ 2 $\times$ 1 k-point grid for
supercell calculations. All atomic coordinates in
supercell containing defect are relaxed using Broyden--Fletcher--Goldfarb--Shanno
(BFGS) quasi--Newton algorithm  until
total energy and forces are converged to $10^{-3}$ eV and $0.025$ eV$/$\AA{} respectively.

The GW \cite{gw_thry1,gw_thry2} calculations are performed using the BerkeleyGW package \cite{BGW}. The number of
unoccupied bands are 100 and 125 for unit cell in monolayer and bilayer respectively.
This choice ensures the convergence of the band gap to be better than 0.1 eV.
The dielectric matrix is expanded in plane wave with energy up to 12 Ry and extended to
finite frequencies using generalized plasmon pole (GPP) \cite{gpp} model. The Coulomb
interaction along the out of-plane-direction is truncated to compute the dielectric
matrix and self--energy \cite{cellslab}. The static remainder technique is
used to accelerate the convergence of the calculation with the number of empty bands
\cite{static_ch}. The Brillouin zone sampling used for GW calculations of unit cell is
21 $\times$ 15 $\times$ 1. We find the quasiparticle band gap for monolayer to be 2.07 eV
which is in good agreement with previous theoretical \cite{bP5,bP6,bP7} and experimental
(scanning tunneling spectroscopy \cite{bp_exp1} and photoluminescence excitation spectroscopy \cite{bp_exp2})
studies. The band gap for bilayer is 1.29 eV which is also consistent
with previous calculations \cite{bP6,bP7}. For the defect calculations with
7 $\times$ 5 supercell in monolayer and bilayer, 3500 and 4300 unoccupied states are used respectively. 
We diagonalize the Hamiltonian using PRIMME software \cite{primme1,primme2} 
to generate the large number of unoccupied bands needed in the calculation.
The Brillouin zone is sampled using a 3 $\times$ 3 $\times$ 1 k-point
grid.

The formation energy of an isolated defect
is defined as:
\begin{align}
 \mathrm{E}^{\mathrm{f}}_{\mathrm{q}}[\vec{\mathrm{R}}_\mathrm{q}](\mathrm{E}_{\mathrm{F}})=\mathrm{E}_{\mathrm{q}}[\vec{\mathrm{R}}_{\mathrm{q}}]-\mathrm{E}_{\mathrm{pristine}}+\mathrm{N}_\mathrm{P}\mathrm{\mu}_{\mathrm{P}}+\mathrm{q}(\mathrm{E}_{\mathrm{F}}+\mathrm{\epsilon}_{\mathrm{VBM}})+\mathrm{E}_{\mathrm{corr}}(\mathrm{q})
\end{align}
where $\mathrm{E}_{\mathrm{pristine}}$ is the total energy per supercell of pristine mono-- or bi--layer
phosphorene,
$\mathrm{E}_{\mathrm{q}}[\vec{\mathrm{R}}_\mathrm{q}]$ is the total energy of a supercell
containing defect in charge state q at its relaxed coordinates $\vec{\mathrm{R}}_\mathrm{q}$, $\mathrm{N}_{\mathrm{P}}$ is
the number of phosphorus atoms added to (or removed from) the supercell to create the
defect, $\mathrm{\mu}_\mathrm{P}$ is the chemical potential of a phosphorus atom which is calculated from 
bulk black phosphorus,
$\mathrm{\epsilon}_{\mathrm{VBM}}$ is the
energy of the valence band maximum (VBM) of pristine cell, $\mathrm{E}_\mathrm{F}$ is the Fermi level with
respect to $\mathrm{\epsilon}_{\mathrm{VBM}}$ 
and $\mathrm{E}_{\mathrm{corr}}(\mathrm{q})$ is the electrostatic correction term.

As discussed previously, simulations of charged defects suffer from an erroneous electrostatic
interaction between the defect cell and its images arising due to periodic boundary conditions. 
As a consequence, formation energies and eigenvalues of defect levels in charged defects are
estimated incorrectly.
The correction term $\mathrm{E}_{\mathrm{corr}}(\mathrm{q})$ in Eqn. (1)
removes the spurious electrostatic interactions and includes the potential alignment
in formation energy.
$\mathrm{E}_{\mathrm{corr}}(\mathrm{q})$ is calculated using the Freysoldt-- Neugebauer--
Van de Walle (FNV) scheme \cite{FNV} as implemented in CoFFEE code \cite{coffee}.
To calculate the correction term,
mono-- and bi--layer phosphorene are modeled as a dielectric slab of width 5.22 \AA{}
and 10.44 \AA{} respectively. The dielectric constants of the slab \cite{ecorr} are calculated
using density functional perturbation theory of the
Quantum Espresso package \cite{qe}.
The in-plane dielectric constants for monolayer
phosphorene are 12.5 and 18.0 along the zigzag direction
and armchair direction, respectively.
The out-of-plane dielectric constant is 1.9. These
quantities for bilayer are 13.5, 27.4 and 1.9 respectively. The difference in the
in-plane dielectric constants is due to the structural anisotropy of phosphorene. The charge
is modeled as Gaussian distribution whose parameters are taken from DFT calculations.
Using this model, the electrostatic energy is calculated for different cell sizes with a uniform
scaling parameter, $\mathrm{\alpha}$, and extrapolated to infinite limit using 
5th order polynomial to obtain the value for an isolated defect \cite{ecorr}.
The difference between the isolated value and the value for a particular cell size gives
the required correction in formation energy for the corresponding cell.
To verify our model,
we calculate the electrostatic energies of negatively charged interstitial defect in bilayer phosphorene with
a model charge for several supercell sizes.
Our starting cell is the 7 $\times$ 5 supercell ($\alpha=7$) for which the in-plane lattice parameters are almost equal.
We uniformly scale the cell size along all the three directions and calculate the electrostatic energy (blue curve in Fig. 1(a)). Further, we perform a
different set of scaling calculations starting from a supercell such that the
vacuum scaling is different. The scaling factor in these cells along the out-of-plane direction 
is 1.5 times of those in
the in-plane directions (red curve in Fig. 1(a)).
We observe that the electrostatic correction does not change monotonically with the inverse of supercell
length which necessitates the several calculations of large supercell to reach the infinite length limit.
As a result, a simple 1/L extrapolation from a small supercell size can lead to
incorrect isolated value \cite{komsa_corr1,komsa_corr2}.
Fig. 1(b) shows the formation energies of negatively charged interstitial in bilayer phosphorene
without and with correction for supercell sizes $7\times5, 10\times8$ and $14\times10$. The corrections have been
obtained following the blue curve of Fig. 1(a). It can be clearly seen that the corrected formation energies
are same for all the three supercells.
For comparison we
have further included the formation energies of neutral defect in bilayer phosphorene for the same
supercell sizes as mentioned above. The formation energies of neutral defect also have no supercell size dependence.
\begin{figure}[]
\centering
\includegraphics[scale=0.80]{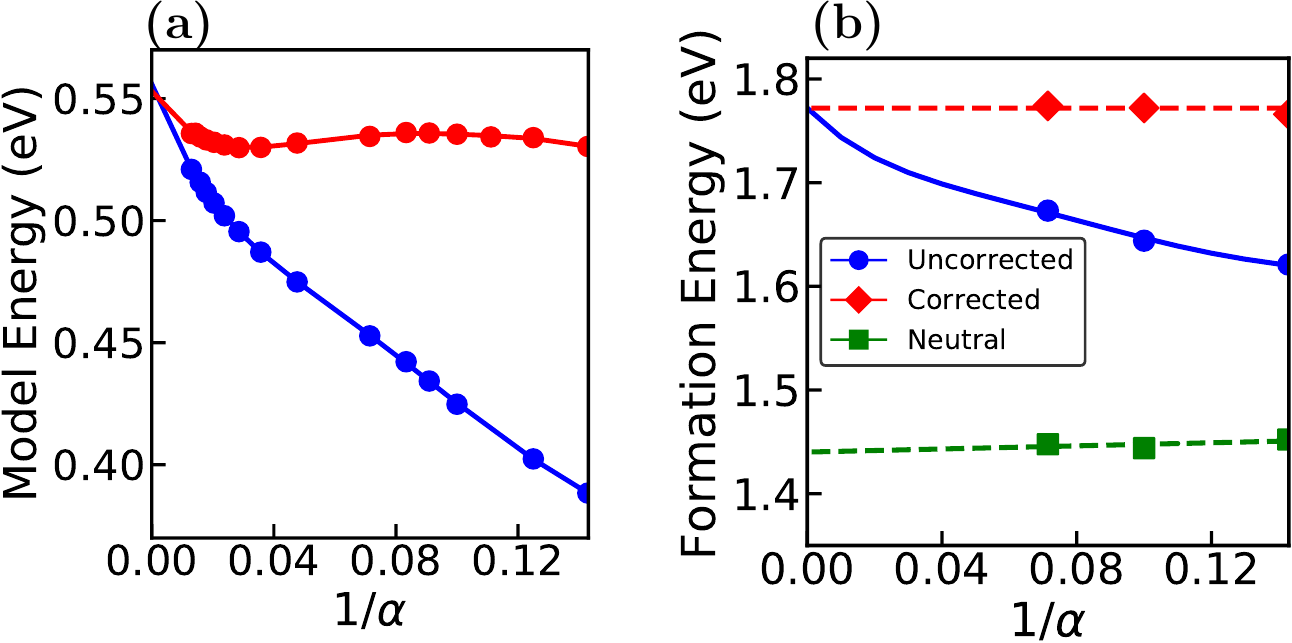}
\caption
{\label{fig0}
(a) shows the variation of electrostatic energy of bilayer phosphorene with different supercell sizes. $\alpha$ is the scaling factor. Blue and red curves represent the variation with scaling $\alpha \times \alpha \times \alpha$ and $\alpha \times \alpha \times 1.5\alpha$ respectively.
In (b) the formation energies of negatively charged interstitial in bilayer phosphorene without and with correction for three different cell sizes are shown by blue circles and red diamonds and extrapolated to infinite limit. The formation energy of neutral interstitial are shown by green squares.
}
\end{figure}

The defect CTL $\mathrm{\varepsilon}(\mathrm{q/q'})$ is defined as the Fermi--level
position for which the formation energies of charge states $\mathrm{q}$ and $\mathrm{q'}$ are equal:
\begin{align}
\mathrm{\varepsilon}(\mathrm{q/q'})
=\frac{\mathrm{E}^\mathrm{f}_{\mathrm{q}}[\vec{\mathrm{R}}_{\mathrm{q}}](E_F=0) -\mathrm{E}^\mathrm{f}_{\mathrm{q'}}[\vec{\mathrm{R}}_{\mathrm{q'}}](E_F=0)}{\mathrm{q'-q}}
\end{align}
The CTL can be computed from formation energies within DFT.
However, due to band gap underestimation within DFT and the fact that CTL involves change in electron number,
 significant error arises in
the computed CTL. Within the combined DFT and GW approach, the CTL
\cite{ctl} is written as:
\begin{align}
\mathrm{\varepsilon}(\mathrm{q/q'}) &= \frac{(\mathrm{E}^\mathrm{f}_{\mathrm{q}}[\vec{\mathrm{R}}_{\mathrm{q}}] - \mathrm{E}^\mathrm{f}_{\mathrm{q'}}[\vec{\mathrm{R}}_{\mathrm{q}}]) + (\mathrm{E}^\mathrm{f}_{\mathrm{q'}}[\vec{\mathrm{R}}_{\mathrm{q}}] - \mathrm{E}^\mathrm{f}_{\mathrm{q'}}[\vec{\mathrm{R}}_{\mathrm{q'}}])}{\mathrm{q'-q}} 
\end{align}
by adding and subtracting term $\mathrm{E}^\mathrm{f}_{\mathrm{q'}}[\vec{\mathrm{R}}_\mathrm{q}]$ in the numerator. 
If the charge states q and q' differ by $\pm 1$, the term $(\mathrm{E}^\mathrm{f}_{\mathrm{q}}[\vec{\mathrm{R}}_{\mathrm{q}}] - \mathrm{E}^\mathrm{f}_{\mathrm{q'}}[\vec{\mathrm{R}}_{\mathrm{q}}])$ can be identified as
the quasiparticle energy ($\mathrm{E}_{\mathrm{QP}}$) which is calculated using GW. This term accounts for
an electron removal or addition to the system.
The other term $(\mathrm{E}^\mathrm{f}_{\mathrm{q'}}[\vec{\mathrm{R}}_{\mathrm{q}}] - \mathrm{E}^\mathrm{f}_{\mathrm{q'}}[\vec{\mathrm{R}}_{\mathrm{q'}}])$ captures the
relaxation energy ($\mathrm{E}_{\mathrm{relax}}$) of the structure due to the change in electron number and is calculated using DFT.
Depending on the addition or removal of the electron (Eqn. 3), CTL can be calculated following different paths (Fig. 2).
The parabolas in Fig. 2 represent the formation energies as a function of the generalized coordinates
of the atoms in the cell.
Along one path, we start with the defect in charge state q at its equilibrium coordinates
, $\vec{\mathrm{R}}_\mathrm{q}$. The $\mathrm{E}_{\mathrm{QP}}$ is represented by the
green vertical arrow, and  $\mathrm{E}_{\mathrm{relax}}$ by the green curved arrow.
The CTL can also be computed starting with the defect in charge state q', as shown by the
red arrows. Since CTL is a thermodynamic quantity, the CTL computed using the two paths should
be the same. Note that if the GW calculation is performed on charged defects the 
defect eigenvalues have to be corrected.
We calculate the CTLs for interstitial in monolayer phosphorene following two paths \cite{gw_ctl}
starting from neutral defect (q=0) and charged defect supercells (q'=+1 or --1). With the corrected quasiparticle values,
the CTLs from different paths agree to within $\pm$ 100 meV. The details of the calculations
are given in subsequent section. 

\begin{figure}[]
\centering
\includegraphics[scale=0.20]{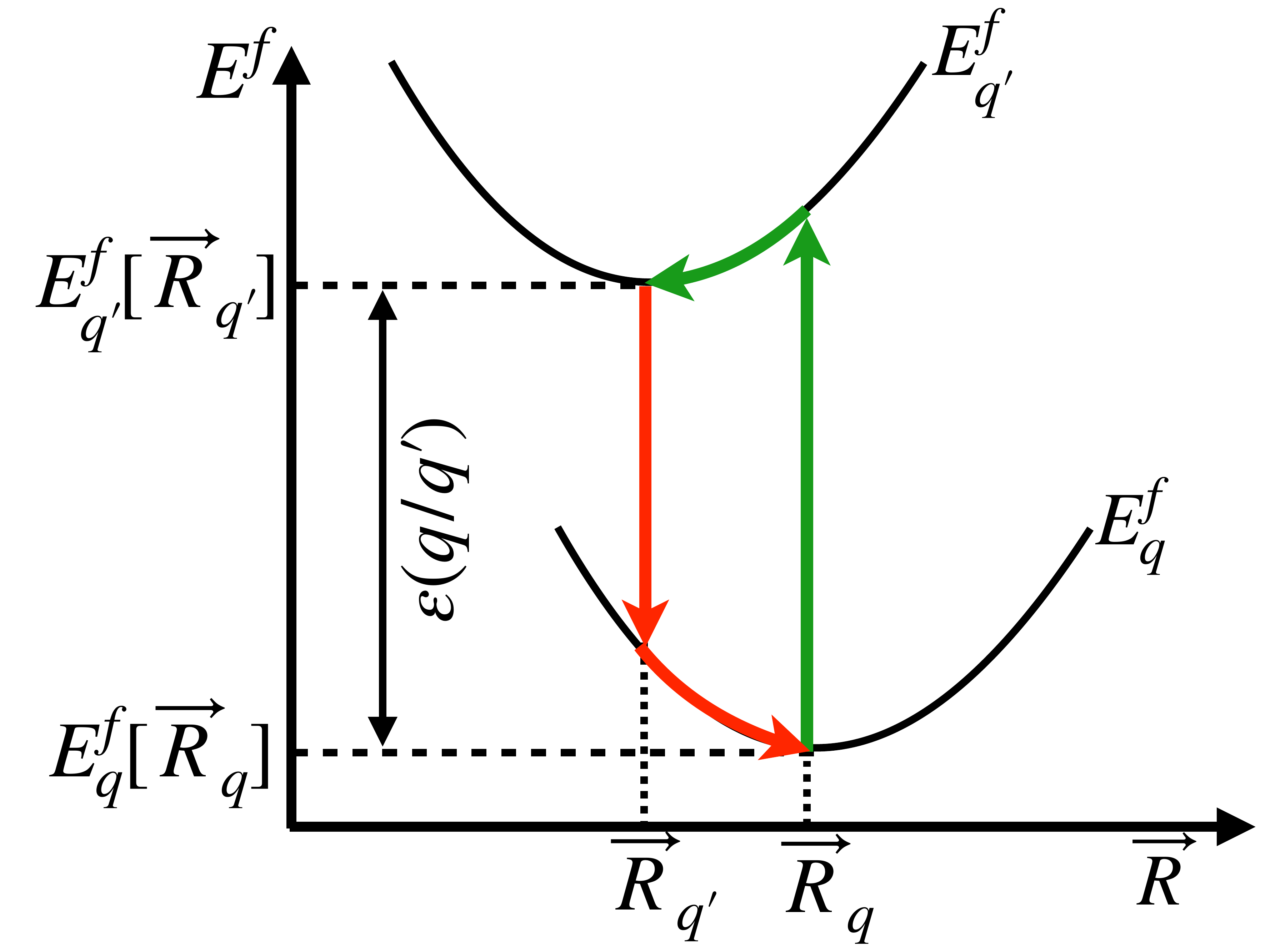}
\caption
{The figure shows calculation of CTLs following different paths. The straight green and red
arrows represent the quasiparticle energies calculated at the equilibrium structures of the
defect with charge state q and q' respectively. The curved arrows account for relaxation
energies.
}
\end{figure}

\section{Defects in Monolayer Phosphorene}

\subsection{Interstitial}

\begin{figure}
\centering
\includegraphics[scale=0.9]{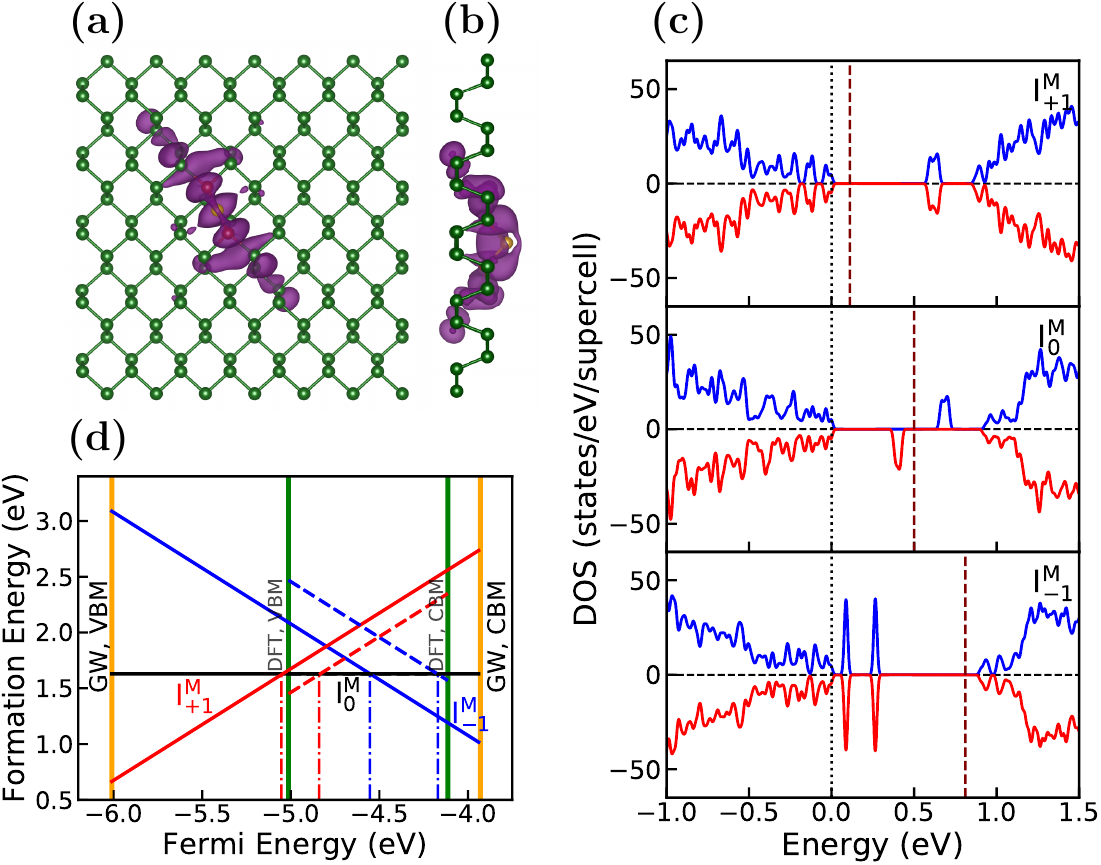}
\caption
{\label{fig2}
(a) Top view and (b) side view of interstitial defect in monolayer phosphorene with defect wave function. The isosurface is plotted at the charge density value of 3.4$\times10^{-3}$ e/\AA{}$^{3}$.
The insterstitial atom is shown in yellow color and the neighbor atoms are in red.
(c) Spin polarized DOS of the interstitial in 3 charged states. VBM is set to zero and is shown in black dotted line. Fermi level is marked in maroon dashed line. Red and blue are for two spin states.
(d) CTLs calculated within DFT and DFT+GW formalism. The dashed and solid lines represent the formation energies
calculated with DFT and DFT+GW respectively and red, black and blue are used for +1, 0 and --1
charge states respectively. Formation energy of neutral defect remains same. The VBM and CBM are also marked in the
figure. We have set the vacuum to zero in the plot.
}
\end{figure}

The most stable structure of interstitial in monolayer phosphorene is when the interstitial atom forms
two symmetric bonds with two phosphorus atoms in one of the sub-layer and resides above the layer (Fig. 3(a) and 3(b)).
Formation of two bonds leaves the interstitial atom with a lone electron which can induce defect states.
Spin--polarized
calculations reveal two defect states in the band gap.
The defect states are localized along one diagonal of the supercell centering the
interstitial atom as is shown in the plot of wave function (Fig. 3(a) and 3(b)). One of the
defect states is occupied in neutral state making interstitial to be stable in all three charge states (+1,0,--1).
Fig 3(c) shows the density of states (DOS)
for all the charge states. Due to the dispersion along the diagonal direction,
the defect bands are broad. 
Upon accepting an electron, interstitial atom breaks the symmetry and the bonds become asymmetric. There is
no significant relaxation when the defect is positively charged. In the negatively charged interstitial, the defect state appears flatter.
This is due to the fact that the defect wave function is more localized.
While the defect states are spin split in the neutral state, they
become degenerate in both positively and negatively charged states as the electrons get paired in these states.
The interstitial defect has low formation energy of 1.63 eV in neutral state.
To obtain the formation energy in charged states,
we evaluate the correction term ($E_{corr}$) 
using CoFFEE as detailed in section 2. The correction using model charge calculation
is 0.295 eV. Taking the potential alignment into accout, $E_{corr}$ is calculated
to be 0.30 eV and 0.26 eV for positively and negatively charged states respectively.
The variation of formation energies of interstitial in charged states, as the Fermi energy is tuned, is shown
in Fig. 3(d). Within DFT (red and blue dashed line),
the defect
changes its state from positive to neutral ($\varepsilon(+1/0)$) at 0.18 eV and from neutral to
negative ($\varepsilon(0/-1)$) at 0.84 eV above VBM (Fig. 3(d)).
As already mentioned, to get a better estimate of CTL we perform GW calculation on the neutral vacancy to
obtain the quasiparticle energy and calculate formation energies within the combined formalism of DFT and GW.
The solid red and blue line in Fig. 3(d) show the variation
of formation energies of positively and negatively charged interstitial respectively within DFT + GW.
The formation energy of neutral defect is same in DFT and DFT + GW.
The DFT + GW CTLs $\varepsilon(+1/0)$ and $\varepsilon(0/-1)$ are at 0.96 eV
and 1.48 eV above VBM respectively. In this DFT + GW calculation, the $E_{QP}$'s are calculated from
the GW calculation of the ionization potential and electron affinity  of neutral interstitial.
CTLs can
be also calculated starting from charged defects (Eqn. 3) \cite{gw_ctl} as described before. 
The CTLs from two paths and their constituent energies are reported in Table 1.
To obtain $\varepsilon(+1/0)$, we
calculate the electron affinity of positively charged interstitial within GW.
In this case, the GW defect level also has to be corrected \cite{gw_ctl}.
The correction for the defect levels is 0.59 eV \cite{coffee}.
Following this path, $\varepsilon(+1/0)$ is calculated to be 1.00 eV. Similarly, starting from
negatively charged interstitial and calculating ionization energy we obtain $\varepsilon(0/-1)$ to be 1.36 eV. The difference
between the CTLs obtained from the neutral and charged cell calculations lie
within the error of GW calculation (0.1 eV).
As a result, the CTLs due to interstitial in monolayer phosphorene are deep in
the gap. The presence of both $\varepsilon(+1/0)$ and $\varepsilon(0/-1)$ CTLs implies that
interstitial in monolayer phosphorene can show both donor-- and acceptor--type behaviors.

\begin{table}[]
\caption{CTLs of interstitial along two different paths are reported. GW calculation on neutral defect
is denoted by path 1 and path 2 represents calculations starting from charged defects.
 All the energies are reported in eV.
E$_{QP}$ is quasiparticle energy and E$_{Relax}$ is the energy associated with relaxing the defect. The
correction for the eigenvalue is given by $\epsilon_{corr}$.}
\begin{tabular}{|c||c|c|c|c||c|c|c|c||c||c|}
\hline
     & \multicolumn{4}{c||}{Path 1}           & \multicolumn{4}{c||}{Path 2}           &    &      \\ \hline
     & E$_{QP}$ & E$_{relax}$ & $\epsilon_{corr}$ & CTL  & E$_{QP}$ & E$_{relax}$ & $\epsilon_{corr}$ & CTL  & Avg. CTL & Difference \\ \hline 
+1/0 & 0.71      & 0.25         & 0.00 & 0.96 & 1.78      &   -0.19      & -0.59& 1.00 & 0.98 & 0.04    \\ \hline
0/-1 & 1.75      & -0.27        & 0.00 & 1.48 & 0.46      &    0.31      &  0.59& 1.36 & 1.42 & 0.12    \\ \hline
\end{tabular}
\end{table}

\subsection{Vacancy:}
Upon removing one phosphorus atom from phosphorene, there are two possible mono--vacancy structures:
MV-(55\textbar66) (Fig. 4(a) and 4(b)) and MV-(5\textbar9) (Fig. 4(c) and 4(d)) \cite{vac1}.
MV-(55\textbar66) has a symmetric structure.
In this structure, the atom closest to the vacancy is connected with 4 P atoms instead of the
usual coordination of 3. In contrast, the closest atom to the vacancy
in MV-(5\textbar9) moves and the system rearranges such that it bonds with 3 atoms. 
The formation energy of MV-(5\textbar9) is lower than that of MV-(55\textbar66) by 330 meV per defect. 
In order to estimate the barrier between the two structures,
we perform a climbing image nudged elastic band (CI-NEB) calculation as implemented in PASTA
package \cite{pasta}. Fig. 4(e) shows how the formation energy changes as
MV-(55\textbar66) transforms to the more stable MV-(5\textbar9) structure. The energy
barrier between the two structures is found to be only 5 meV. 
Due to this low barrier, we expect the defect to always be in the MV-(5\textbar9) structure.
Hence, for all further calculations, we only consider
the more stable structure, MV-(5\textbar9).

\begin{figure}
\centering
\includegraphics[scale=0.8]{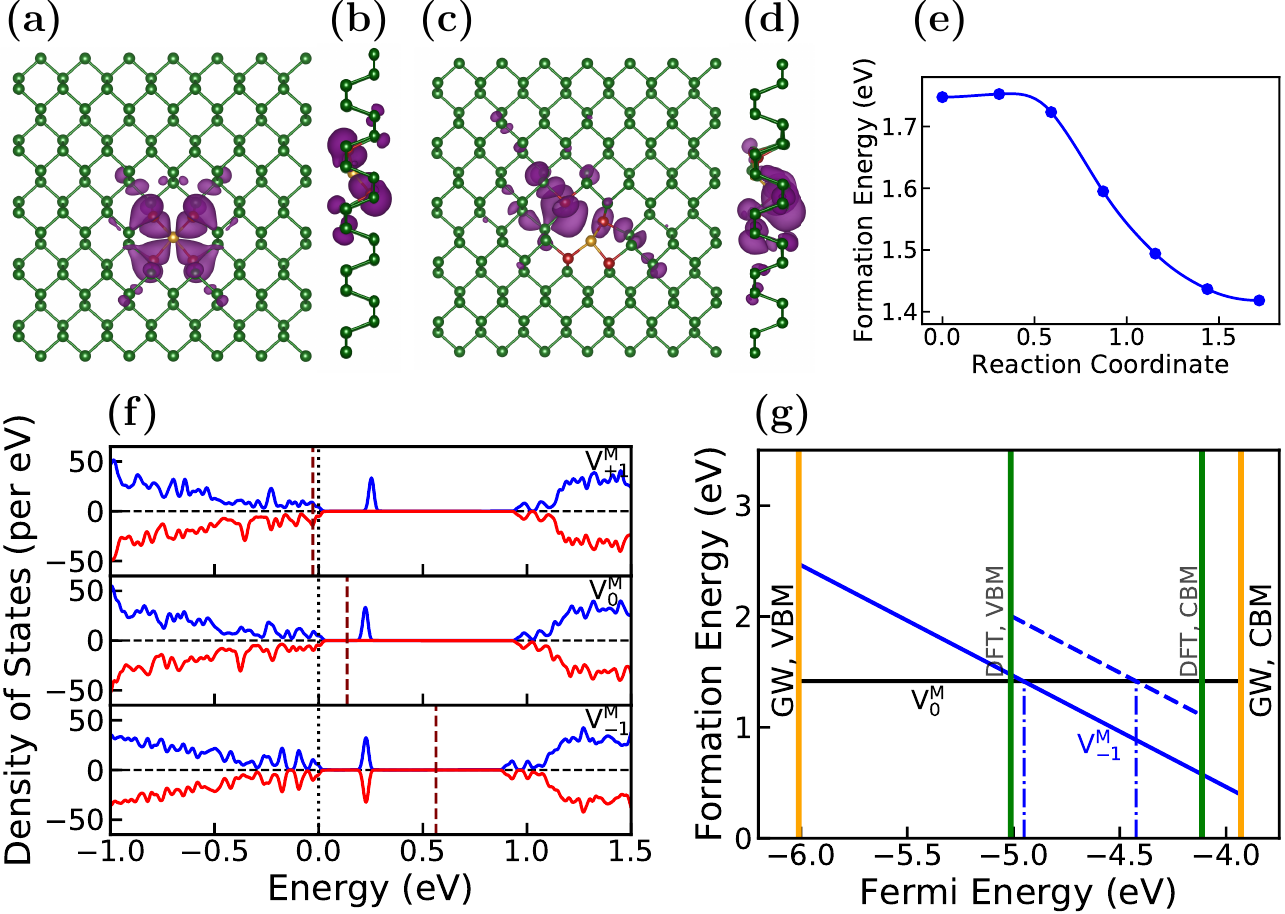}
\caption
{\label{fig1}
(a) and (b) show the top and side view of MV-(55\textbar66) in 7$\times$5 supercell and defect wave function.
(c) and (d) show  MV-(5\textbar9) and the  wave function localized at defect from top and side
view respectively. The isosurface of wave functions is plotted at 3.4$\times10^{-3}$ e/\AA{}$^{3}$.
The neighbour to the vacancy is shown in yellow and the other neighbour atoms are shown in red.
(e) shows the variation of formation energy while the structure changes from MV-(55\textbar66)
to MV-(5\textbar9). The reaction coordinate is the distance between two images in hyper-surface.
(f) Spin--polarized DOS of vacancy in all 3 charge states. The red and blue lines denote two spin states. The VBM of the corresponding cells
are set to zero and shown in black dotted lines. The maroon dashed lines are Fermi levels of
the corresponding systems.
(g) depicts the variation of formation energies with Fermi energy both within DFT and DFT+GW. The
black line is for vacancy in neutral state and negatively charged vacancy is represented by the
blue solid (DFT+GW) and dashed (DFT) lines.
}
\end{figure}

We investigate the MV-(5\textbar9) vacancy in 3 charge states: neutral, positive and negative.
We perform spin--polarized calculations. The neutral vacancy gives rise to an unoccupied flat band
in the band gap. The DOS calculation (Fig. 4(f)) shows a state 0.23 eV above the VBM.
We plot the
corresponding wave function in Fig. 4(c) and 4(d). It is clear that the wave function of this site is localized around
the vacancy along a diagonal of the supercell. We can understand the origin of this state as follows:
a P atom neighbouring the vacancy forms two bonds and has a
dangling bond which gives rise to the flat band in the gap.
The filled defect state lies deep within the valence band and
we found that the vacancy is not stable in positive charge state. 
The defect can accommodate an electron and change its state to negatively
charged state. Upon accepting an electron, the system rearranges itself such that
both the defect bands are in the band gap and they become degenerate
(Fig. 4(f)). It is to be noted that MV-(55\textbar66) also induces defect states but those are
within the valence band. These states hybridize with the valence band states making the defect always negatively charged.
However, this configuration is always higher in energy than negatively charged MV-(5\textbar9).

We calculate the formation energies of MV-(5\textbar9) in neutral and negatively charged states.
In Fig. 4(g)) the black line represents the formation energy of neutral vacancy. 
The vacancy has a formation energy of 1.42 eV in the neutral state.
Due to this low formation energy, vacancy defects are expected to be abundant in phosphorene \cite{bp_expt}.
The variation of formation energy in negatively charged state with the Fermi energy is shown in Fig. 4(g) within DFT
(blue dashed line) and DFT+GW (blue solid line).
The CTL $\varepsilon(0/-1)$ is 0.62 eV above the VBM within DFT and 1.06 eV above
the VBM within DFT + GW.
This implies that vacancy in monolayer phosphorene behaves as deep acceptor.

\section{Defects in Bilayer Phosphorene}

\subsection{Interstitial}

\begin{figure}
\centering
\includegraphics[scale=0.9]{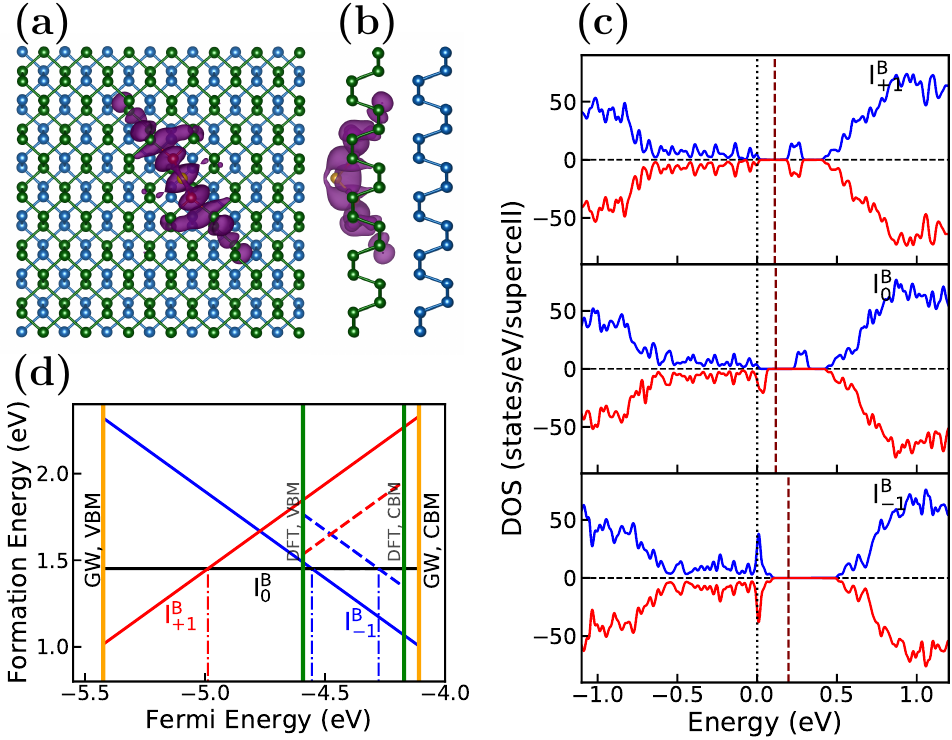}
\caption
{
\label{fig4}
(a) and (b) are the top and side view of interstitial defect with the localized wave function in
bilayer phosphorene.
(c) Spin--polarized DOS of interstitial in bilayer phosphorene in positive, neutral and negative charged states. VBM are set to zero and the Fermi level are shown in maroon dashed line. Red and
blue lines represent two spin states.
(d) CTLs calculated within both DFT and DFT+GW. Solid lines denote the formation energies within
DFT+GW and the dashed lines are used for DFT. Red, black and blue are for +1, 0 and --1 charge states. Vacuum is set to zero.
}
\end{figure}

Interstitial in bilayer has a similar structure to that of interstitial in monolayer phosphorene
(Fig. 5(a) and 5(b)).
The interstitial atom prefers to bond with two atoms in the outer sub-layer of bilayer phosphorene facing vacuum.
The configuration with the interstitial atom between the layers is higher in energy by 580 meV per defect.
Like the monolayer, spin--polarized calculation on interstitial in bilayer shows that it also induces two
defect states in the gap. One of the defect states is filled rendering interstitial to be possible
in +1, 0, --1 charge states. DOS calculations on the three charge states reveal that while the
defect states are non-degenerate in neutral state they become degenerate by accepting or donating
an electron in charged states (Fig. 5(c)). 
In the neutral charge state, the formation energy is 1.45 eV which is lower than the corresponding defect
in the monolayer.
Fig. 5(d) shows the formation energies of interstitial in its three charge states +1,
0 and --1 within both DFT and DFT+GW.
We find the +1 charged interstitial is always higher in energy when the Fermi energy is in band
gap in DFT. In contrast, the interstitial is more stable in positively charged state within DFT+GW when
Fermi level is below 0.43 eV with respect to VBM which marks the CTL $\varepsilon(+1/0)$.
The CTL $\varepsilon(0/-1)$ is at 0.86 eV (0.33 eV) above VBM within DFT+GW (DFT).
This suggests that the interstitial in bilayer can act as both donor-- and acceptor--type defect like
the interstitial defect in monolayer.

\subsection{Vacancy}

\begin{figure}[h]
\centering
\includegraphics[scale=1.2]{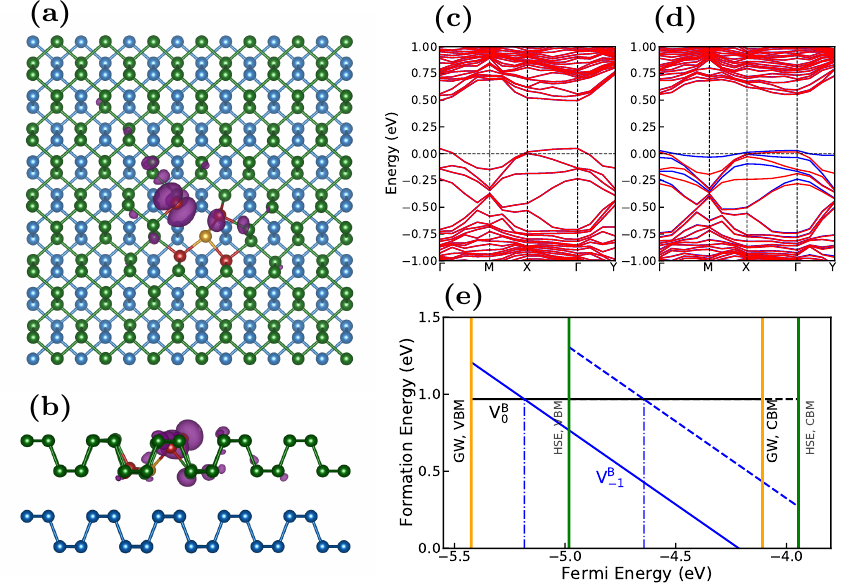}
\caption
{
\label{fig3}
(a) and (b) show the top and side view of MV-(5\textbar9) in bilayer phosphorene. The defect
wave function is plotted at isosurface value of 3.4$\times10^{-3}$ e/\AA{}$^{3}$.
(c) and (d) show the band diagrams of the defect structure relaxed with PBE and HSE respectively.
Red and blue lines denote two spin states.
(e) is the formation energy plot of vacancy in neutral and negative state.
}
\end{figure}

In bilayer phosphorene, the vacancy can reside in two inequivalent positions.
In one configuration the vacancy faces the second layer while in the other it
faces vacuum.
We found the vacancy is most stable in MV-(5\textbar9) structure facing vacuum (Fig. 6(a) and 6(b)).
Spin--polarized calculation within GGA-PBE on this structure does
not show any distinct defect state in the band gap (Fig. 6(c)).
This calculation suggests hybridization between defect state and states inside valence band edge.
However, after a diagonal GW correction, a state emerges in the gap.
This poses a problem to the diagonal GW approximation (Eqn. 4), in which the self--energy matrix element is
computed using the DFT wave functions as:
\begin{equation}
\mathrm{E}^{\mathrm{QP}}_{\mathrm{i}} = \mathrm{E}^{\mathrm{DFT}}_{\mathrm{i}} + \langle\mathrm{\psi}^{\mathrm{DFT}}_\mathrm{i}\lvert\mathrm{\Sigma}_{\mathrm{GW}}(\{\mathrm{E}^{\mathrm{DFT}},\mathrm{\psi}^{\mathrm{DFT}}\};\mathrm{E}^{\mathrm{QP}})\lvert\mathrm{\psi}^{\mathrm{DFT}}_\mathrm{i}\rangle - \langle\mathrm{\psi}^{\mathrm{DFT}}_\mathrm{i}\lvert \mathrm{V}_{\mathrm{XC}}\lvert\mathrm{\psi}^{\mathrm{DFT}}_\mathrm{i}\rangle
\end{equation}
One way to overcome this problem is to perform self--consistent GW calculation. Another way is
to choose a better mean--field. To address the issue we start with alternative mean--field. We use
hybrid functional approximation (HSE) \cite{hse} for exchange--correlation which is an improvement over GGA. 
We relax the vacancy structure with HSE. For all further calculations, the total energies are calculated
with HSE.
The obtained band structure gives a distinct state in the band gap, however the state is very close to
valence bands (Fig. 6(d)).
Vacancy has a formation energy of 0.97 eV in neutral state.
The calculated CTL $\varepsilon(0/-1)$ within HSE is at 0.34 above VBM (Fig. 6(e)).
For GW calculation we choose a configuration such that the defect state is in the gap within HSE. As
generating 4300 bands with HSE is computationally expensive and the quasiparticle energies depend weakly
on the mean--field used, we generate the wave functions within PBE. 
The starting PBE defect wave function is localized at the defect site.
While the defect state is also in the gap using PBE, because of its proximity to the VBM edge
it is expected
that the diagonal $G_0W_0$ approximation is not going to be adequate. 
We construct the GW Hamiltonian in the DFT wave function basis
following Eqn. 5.
\begin{equation}
\mathrm{H}_{\mathrm{ij}} = \mathrm{E}^{\mathrm{DFT}}_{\mathrm{i}}\mathrm{\delta}_{\mathrm{ij}} + \langle\mathrm{\psi}^{\mathrm{DFT}}_\mathrm{i}\lvert\mathrm{\Sigma}_{\mathrm{GW}}(\{\mathrm{E}^{\mathrm{DFT}},\mathrm{\psi}^{\mathrm{DFT}}\};\mathrm{E}^{\mathrm{QP}})\lvert\mathrm{\psi}^{\mathrm{DFT}}_\mathrm{j}\rangle - \langle\mathrm{\psi}^{\mathrm{DFT}}_\mathrm{i}\lvert \mathrm{V}_{\mathrm{XC}}\lvert\mathrm{\psi}^{\mathrm{DFT}}_\mathrm{j}\rangle
\end{equation}
Due to computational cost, we restrict the construction of Hamiltonian matrix with wave functions with energy within $\pm$500 meV of the defect state as we are interested in the defect state and the states close to it.
The self--energy matrix is evaluated within static--screening limit (static--COHSEX) \cite{gw_thry2} and
we diagonalize the constructed Hamiltonian \cite{sc_gw}.
The eigenfunctions of the Hamiltonian are the new wave functions with which we performed $G_1W_0$ calculation.
It should be noted that in this way, the wave functions are iterated while the dielectric screening is still
constructed from mean--field wave functions.
The quasiparticle energies are obtained by evaluating the self--energy using the standard plasmon pole model.
For the vacancy in bilayer, $G_0W_0$ calculation shows that there is mixing between the defect state and the valence bands close to it. 
After the first iteration with the updated wave functions, the quasiparticle energy of the defect state
is 370 meV away from the VBM edge.
 
Instead of calculating the CTL using the quasiparticle energy at the equilibrium structure, we can calculate
that at any  structure and adjust the relaxation energy accordingly (Fig. 7) \cite{gw_path}. We can denote an intermediate
structure by $\vec{\mathrm{R}}_\mathrm{I}$ and rewrite the Eqn. (2) in following way:
\begin{align}
\mathrm{\varepsilon}(\mathrm{q/q'}) &= \frac{(\mathrm{E}^\mathrm{f}_{\mathrm{q}}[\vec{\mathrm{R}}_{\mathrm{q}}] - \mathrm{E}^\mathrm{f}_{\mathrm{q}}[\vec{\mathrm{R}}_{\mathrm{I}}]) + (\mathrm{E}^\mathrm{f}_{\mathrm{q}}[\vec{\mathrm{R}}_{\mathrm{I}}] - \mathrm{E}^\mathrm{f}_{\mathrm{q'}}[\vec{\mathrm{R}}_{\mathrm{I}}]) + (\mathrm{E}^\mathrm{f}_{\mathrm{q'}}[\vec{\mathrm{R}}_{\mathrm{I}}] - \mathrm{E}^\mathrm{f}_{\mathrm{q'}}[\vec{\mathrm{R}}_{\mathrm{q'}}])}{\mathrm{q'-q}} \\
&=\mathrm{E}_{\mathrm{relax,q}} + \mathrm{E}_{\mathrm{QP}} + \mathrm{E}_{\mathrm{relax,q'}}
\end{align}

\begin{figure}[]
\centering
\includegraphics[scale=0.20]{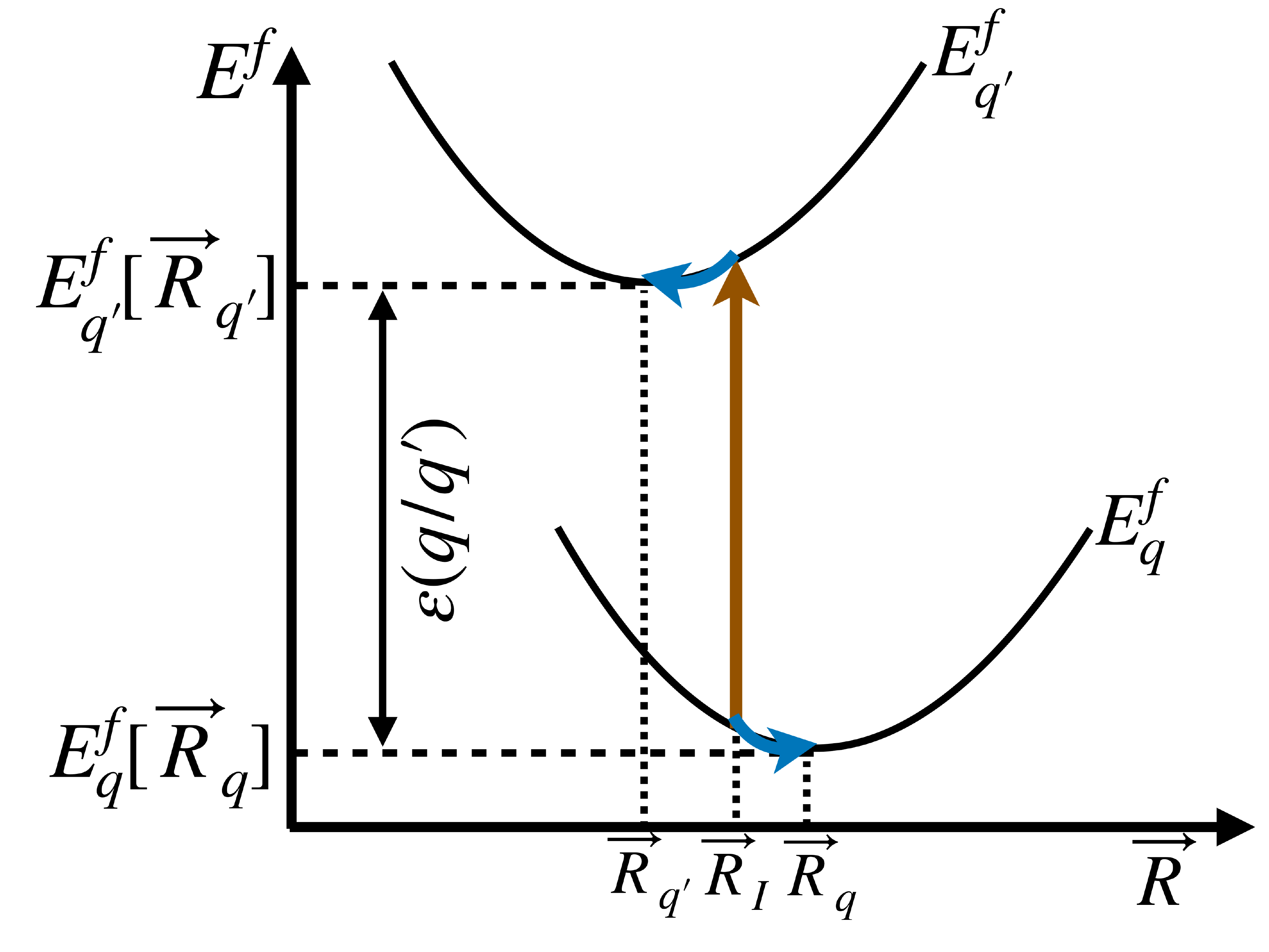}
\caption
{The figure shows calculation of CTL starting from an intermediate structure. The brown
arrow represents the quasiparticle energy calculated at $\vec{\mathrm{R}}_\mathrm{I}$
and the curved blue arrows are the relaxation energy.
}
\end{figure}

Fig. 7 shows two energy curves for two charge states of the defect. $\vec{\mathrm{R}}_\mathrm{I}$ is an intermediate
structure at which the quasiparticle energy is calculated (brown straight arrow in Fig. 7). To obtain the CTL,
the relaxation energies from the $\vec{\mathrm{R}}_\mathrm{I}$ to $\vec{\mathrm{R}}_\mathrm{q}$ and $\vec{\mathrm{R}}_\mathrm{q'}$ (blue curved
arrows in Fig. 7) are taken into account.
For the vacancy calculation in bilayer phosphorene, we adopt this scheme to calculate the CTL within DFT + GW.
The intermediate structure is chosen as discussed above.
We perform the GW calculation with the intermediate structure to get the quasiparticle energy
($\mathrm{E}_{\mathrm{QP}}$) and calculate the relaxation energy from that structure to equilibrium neutral
and negatively charged structure as the vacancy is stable in neutral and negative state.
Using this formalism (Eqn. (6)), the CTL
$\varepsilon(0/-1)$ is found to be at 0.24 eV with respect to VBM within DFT + GW (Fig. 6(e)) which again implies that vacancy in bilayer phosphorene behaves like an acceptor.

\section{Conclusion}
We have extensively studied the formation and electronic properties of vacancy and self--interstitial defects in mono-- and bi--layer phosphorene. 
We have taken into account the spurious electrostatic correction while studying
charged defects.
The defects have formation energies between 0.9 eV -- 1.6 eV in
neutral state. Depending on the charge state, these defects can further lower their formation
energies. It has been also observed that the formation energy of defects in
bilayer phosphorene is smaller than that in monolayer suggesting that with
the increase of layer number formation energy decreases.
We calculate the CTLs of the defects within DFT and DFT+GW formalism which suggest that while
vacancy behaves as acceptor type defect, interstitial can act as both acceptor and donor type defect.

\begin{acknowledgments}
We thank the Supercomputer Education and Research Centre (SERC) at IISc
for providing the computational resources. 
\end{acknowledgments}

%

\end{document}